\documentstyle[seceq,epsf]{ptptex}
\title{%        %You can use \\ for explicit line-break
The Role of Domain Walls on the Vortex Creep Dynamics in
Unconventional Superconductors}
\author{%       %Use \sc for the family name
Manfred {\sc Sigrist} and Daniel F. {\sc Agterberg}$^*$
}
\inst{%         %Affiliation, neglected when [addenda] or [errata]
Yukawa Institute for Theoretical Physics, 
Kyoto University, Kyoto 606-8502, Japan \\
$^*$ National High Magnetic Field Laboratory, Florida State
University, Tallahassee, FL 32306, USA}

\recdate{%      %Editorial Office will fill in this.
\today
}

\abst{%       %this abstract is neglected when [addenda] or [errata]
We investigate the influence of domain walls 
on the vortex dynamics in superconductors with multi-component order
parameters. We show that, due to their complex structure domain walls 
can carry vortices with fractional flux quanta. The decay of
conventional vortices into fractional ones on domain walls is
examined. This decay presents an extraordinarily strong pinning
mechanism for vortices and turns domain walls occupied with pinned fractional 
vortices into efficient barriers for the vortex motion. Therefore,
domain walls can act as fences for the flux flow, preventing the decay
of the remnant magnetic flux enclosed by them. Furthermore, the
consequences of this property of domain 
walls on the vortex dynamics are discussed in connection 
with observed noise in the hysteresis cycle, using the Bean
model of the critical vortex state. Based on this picture experimental
data in the unconventional 
superconductors UPt$_3$, U$_{1-x}$Th$_x$Be$_{13}$ and Sr$_2$RuO$_4$
are interpreted.}

\begin{document}

\maketitle

\section{Introduction}

The penetration of magnetic fields into type-II superconductors as
flux lines (vortices) yields many complex phenomena. In recent years the
physics of vortices has become an important subject, particularly in
connection with high-temperature superconductivity. \cite{BLATTER}
The physics of ``vortex matter'' is crucial for a large number of
applications of superconductivity involving high currents and
fields. Vortices limit the technological capability of
superconductors, since their motion generates dissipation. Therefore 
one of the central issues is the pinning of vortices at defects such
as impurities, lattice dislocations and twin boundaries. The
critical current, the limit of dissipation-free transport, 
depends on the character and strength of the pinning potentials
created by these defects. The effect of pinning was shown to
crucially depend on the various phases of vortex matter, and the
resulting problems are naturally very complex. \cite{BLATTER}

In this paper we would like to discuss a pinning phenomenon in
unconventional superconductors which has a character different from
the usual pinning at crystal defects. This study is motivated by
recent experiments on heavy fermion compounds UPt$_3$ and
U$_{1-x}$Th$_x$Be$_{13}$ as well as on the transition metal 
oxide Sr$_2$RuO$_4$ \cite{AMANN,LT,MOTA,SHUNG}. All these
superconductors have  a comparatively low 
transition temperature of the order of 1K. Fluctuation effects do not
play an important role in these systems. However, interesting vortex
physics is introduced because the superconducting order
parameter of these systems has more than one component.  This yields more
degrees of freedom in forming topologically stable defects of the
order parameter, a well-known feature in superfluid $^3$He physics. We would
first like to review some of the basic experimental facts concerning the
mixed state of these three superconductors, which will be relevant for 
our theoretical discussion.

One aspect of the mixed state related to vortex pinning  is the slow 
motion of vortices close to the critical state, known as flux
creep. This phenomenon is observed as the  
slow decrease of the remnant magnetization after the
superconductor has been exposed for some time to a magnetic field 
considerably larger than the lower critical field, $ H_{c1} $. The remnant
magnetization that exists after turning off the external field 
originates from the
pinning of vortices in the material. The motion of these vortices out
of the sample, generally governed by thermally activated
crossing of pinning potential barriers, leads to the slow decay of the 
magnetization.\cite{BLATTER}  (Note that quantum
phenomena, macroscopic quantum tunneling, can also play an important role
at sufficiently low temperature. \cite{BLATTER,ANACELIA}) The decay
(creep) rate is determined by the 
temperature, the pinning properties, and the vortex matter state as
described by Kim and Anderson \cite{KIM,TINKHAM} (see also Geshkenbein
and Larkin 
\cite{GESH}). A simple and intuitive theory of the
critical vortex state in a material with strong pinning effects
was given by Bean.\cite{BEAN}  It describes the characteristic profile of
the vortex distribution depending on the history of the external
applied fields and the critical current. \cite{TINKHAM}

In the two heavy Fermion superconductors UPt$_3$ and
U$_{0.9725}$Th$_{0.0275}$Be$_{13}$  ($ 0.017 \leq x \leq 0.45 $),
an anomalous temperature dependence of the creep rate was observed by Mota's
group. \cite{AMANN,LT}  UPt$_3$ and 
U$_{1-x}$Th$_x$Be$_{13}$ ($ 0.017 \leq x \leq 0.45 $) both exhibit two
consecutive superconducting phase transitions. Flux relaxation
measurements show a rapid drop of the creep rate, essentially to zero, 
immediately below the second transition in both system.\cite{AMANN,LT} 
A similar sudden drop 
in the creep rate was reported for Sr$_2$RuO$_4$ at the rather low
temperature $ T^* \approx 50$  mK. \cite{MOTA} In contrast to the
former two examples, 
no sign of an additional transition at this temperature is observed 
in any other properties to this time. The transition in the creep rate
indicates the onset of a new efficient pinning mechanism which inhibits 
the motion of vortices from the interior to the surface of the sample. 
However, it
has been noticed in experiments that   
after a long waiting period some creep recovers. \cite{AMANN,MOTA}

This effect is readily explained if we assume that fence-like
structures exist which prevent the passage of vortices
(so that the vortices cannot leave the sample).
Experiments indicate that  
these fences are activated below a certain transition temperature within the
superconducting phase. The effect of the fences is 
rather different
to standard pinning. In particular, their
influence on the critical current is weak, as discussed below.
This is consistent with hysteresis measurements on UPt$_3$ in a slowly
oscillating magnetic field. \cite{SHUNG} While the flux creep 
disappears 
in the low-temperature phase, the critical current is continuously
increasing and shows no anomaly upon entering 
the low temperature phase. \cite{AMANN,LT} Furthermore, 
below the second
transition Rosenbaum's group found that 
the magnetization curves of the hysteresis cycle display strong
variations from cycle to cycle in the region close to $H_{c1} $, while
in the high field part of the cycle (connected with the critical current
of the superconductor) no unusual behavior is seen \cite{SHUNG}. 
This ``noise'' can be understood as being induced by the weakly mobile fences
inhibiting the free flow of vortices. In every cycle the position of the
fences changes slightly, and the magnetization process for small
vortex concentrations, i.e. for fields close to $ H_{c1} $, is
modified. \cite{SHUNG} We will show that a simulation with a simple
model incorporating this effect gives rise to the features 
observed in these experiments.

What is the origin of these fence-like structures? 
In unconventional superconductors
such structures occur if the superconducting phase is degenerate, 
e.g., for time-reversal symmetry breaking states which have at least
a two-fold degeneracy. Degenerate states can appear as domains in the
superconductor accompanied by domain walls as topological structures of
the order parameter. \cite{REVIEW} Various properties of such 
domain walls have been 
studied by several groups. \cite{REVIEW,VOLOVIK1,DOMAIN} Domain walls
in time-reversal symmetry breaking superconductors possess interesting
magnetic properties and carry chiral quasiparticle bound
states.\cite{VOLOVIK1,DOMAIN,MATSUMOTO}  
The question arises whether or not they  serve as barriers to the vortex
motion. The domain wall is naturally accompanied by a slight
local suppression of the order parameter which tends to trap
vortices. In this case the pinning strength is rather weak and should not
have much influence on the flux motion. However, strong pinning can
arise from properties of the internal structure of domain walls.
Under certain circumstances the domain wall has, in addition to  
a stable structure,    
metastable structures, or even two degenerate stable
states. \cite{REVIEW}  These states may form domains on the
domain wall that are separated by line defects. It has
been shown that such lines carry a magnetic flux which 
is an arbitrary fraction of a standard flux quantum in a
superconductor.\cite{DOMAIN} We will show that an ordinary vortex
placed on such a 
domain wall can decay into two fractional vortices. 
\cite{DOMAIN,VOLOVIK2}  Since these fractional 
vortices can only exist on the domain wall, they are strongly
pinned. They repel other approaching vortices, and in this way the
domain wall indeed acts as a strong barrier.  
A high density of vortices close to the domain wall
can destroy the state which carries fractional vortices, and the
pinning effect disappears. This is equivalent to having a barrier 
``height'' for the vortex pinning. It is important to note that this pinning
property of the domain wall does not alter the bulk pinning due to
impurities within each domain. This type of impurity-based pinning is
responsible for determining the critical
current. However, the domain walls play an important role in the
overall flux motion. They are the origin of an {\it intrinsic} pinning
location created by the superconducting state itself,
in contrast to the {\it extrinsic} pinning due to material defects. 
Note, however, that the domain walls themselves are pinned at lattice
defects reducing their mobility.

In this article we first discuss properties of domain walls
in the most simple case of a time-reversal symmetry breaking
superconducting state using the Ginzburg-Landau theory. Then we
use the domain walls as barriers in a model based on the Bean's theory
and consider the modification of flux motion properties due to these barriers.

\section{Properties of a domain wall}

In this section we investigate the structure of a domain wall in a
time-reversal symmetry breaking superconducting state as may be
realized in the low-temperature phases of UPt$_3$ and 
U$_{1-x}$Th$_x$Be$_{13}$, and at the onset of 
superconductivity in Sr$_2$RuO$_4$. 
We do not include the aspects of the double
transition related to the former two, as it complicates the discussion
considerably without leading to further insight. 

\subsection{Ginzburg-Landau theory}

For the following discussion we introduce the simplest
possible model that contains all the relevant features in order to
discuss the physics of domain walls in an unconventional
superconductor. We consider a system with tetragonal crystal symmetry
based on the point group $ D_{4h} $. The superconducting order parameter
shall belong to the two-dimensional representation $ E_g $
(even parity) or $ E_u $ (odd parity), yielding the following
expansion for the gap functions ($2 \times 2 $ gap matrix with the
notation $ \hat{\Delta} = i \hat{\sigma}_y \psi $ for even and $
\hat{\Delta} = i (\mbox{\boldmath $ d $} \cdot \hat{\mbox{\boldmath $
\sigma $}}) 
\hat{\sigma}_y $ for odd parity): 

\begin{equation} \begin{array}{l}
\psi(\mbox{\boldmath $ k $}) = (\eta_x v_z v_x + \eta_y v_z v_y)/\langle v_x^2
v_z^2\rangle , \\ \\ 
\mbox{\boldmath $ d $} (\mbox{\boldmath $ k $}) = \hat{{\bf z}} (\eta_x v_x + \eta_y v_y)/\langle v_x^2 
\rangle . \\
\end{array} \end{equation}
Here $ \mbox{\boldmath $ \eta $} = (\eta_x , \eta_y )$ represent the
two-dimensional complex order parameter, $ \hat{{\bf z}} $ denotes
the unit vector along $ z $-direction, $v_i$ denotes the components of
the Fermi velocity, and $\langle ... \rangle$ is the average over the
Fermi surface.  
As mentioned above, we concentrate on the time-reversal symmetry
breaking states of the form $ \mbox{\boldmath $ \eta $} \propto (1,
\pm i) $ yielding the gap functions

\begin{equation}
\psi (\mbox{\boldmath $ k $}) = \eta_0 v_z (v_x \pm i v_y)/\langle v_x^2 v_z^2
\rangle  \qquad {\rm or } \qquad  
\mbox{\boldmath $ d $} (\mbox{\boldmath $ k $}) = \hat{{\bf z}} (v_x \pm i v_y )/ \langle v_x^2 \rangle.
\end{equation}
It is convenient to transform the order parameter into the 
basis set $ \eta_{\pm} = (\eta_x \mp i \eta_y ) / \sqrt{2} $. 
For simplicity we also use the phenomenological parameters as determined
by weak coupling theory and assume that $v_z$ does not depend
upon $k_x$ or $k_y$. The general GL free energy then has the form

\begin{equation} \begin{array}{ll} \displaystyle
{\cal F} = \int d^3 r & [ \displaystyle a (|\eta_+|^2 + |\eta_-|^2) + b
\{ (|\eta_+|^4 + 
|\eta_-|^4)  + 4 |\eta_+|^2 |\eta_-|^2  \\ & \\
& \displaystyle + \nu (\eta_-^{*2} \eta_+^2 + 
\eta_-^{2} \eta_+^{*2}) \}
+ \kappa \{ |\mbox{\boldmath $ D $} \eta_+|^2 + |\mbox{\boldmath $ D $} \eta_-|^2 
\\& \\ &  \displaystyle +\frac{1}{2}( (D_- \eta_+)^* (D_+ \eta_-) +
\nu (D_+ \eta_+)^* (D_- \eta_-) + {\rm c.c.}) \}
+ \frac{1}{8\pi}( \nabla \times \mbox{\boldmath $ A $})^2 ].
\\& 
\end{array} \end{equation}
Here $ a=a_0 (T-T_c) $ and $b$ and $ \kappa $ are the standard coefficients
derived from the weak coupling theory.
In the gradient terms, we use the gauge 
invariant derivatives $ \mbox{\boldmath $ D $} = \mbox{\boldmath $
\nabla $} - i \gamma
\mbox{\boldmath $ A $} $, with  
$ D_{\pm} = D_x \pm i D_y $ and $
\gamma = 2 e / \hbar c = 2 \pi / \Phi_0 $ ($ \Phi_0 $ is the flux quantum). 
The parameter $ \nu $ denotes the deviation of the Fermi surface or
its density of states from cylindrical symmetry around the $z$-axis due
to the tetragonal crystal field

\begin{equation}
\nu = \frac{ \langle v_{x}^4 \rangle - 3 \langle v_{x}^2
v_{y}^2 \rangle }{\langle v_{x}^4 \rangle+ \langle v_{x}^2
v_{y}^2 \rangle},
\end{equation}
where $ \nu = 0
$ for a cylindrical symmetric Fermi surface.
In this form the GL theory describes two degenerate superconducting
states, $ (\eta_+ , \eta_- ) = \eta_0 (T) (1,0) $ and $ \eta_0 (T)
(0,1) $ with

\begin{equation}
\eta_0^2 (T) = \frac{a_0 (T_c - T)}{2 b}.
\end{equation}
for $ T < T_c $.

In the following we analyze the properties of domain walls
between the two degenerate states. We assume that they are
infinite and planar such that we can separate the coordinates into
components parallel and perpendicular to the normal vector $
\mbox{\boldmath $ n $} $  
of the domain wall. For
simplicity we ignore the $z$-direction, assuming that $
\mbox{\boldmath $ n $} \perp 
{\bf z}$. Therefore it is convenient to 
rewrite the free energy in these coordinates: $ (x,y) \to (x',y')=
(r_{\parallel}, r_{\perp} )$ and $ D_{\pm} = e^{\pm i \theta}
D'_{\pm} $. Simultaneously we transform the order parameter $
\eta_{\pm} = e^{\pm i \theta} \eta'_{\pm} $ so that the gradient terms 
become

\begin{equation}
\kappa \{ |\mbox{\boldmath $ D $}'\eta'_+|^2 + |\mbox{\boldmath $ D $}' \eta'_-|^2 +\frac{1}{2} 
((D'_- \eta'_+)^* (D'_+ \eta'_-) + \nu e^{-i4 \theta}
(D'_+ \eta'_+)^* (D'_- \eta'_-) + c. c.) \}
\end{equation}
and in the homogeneous part only the third term among the fourth order
terms is modified to 

\begin{equation}
b \nu e^{i4 \theta}  \eta_-'^{*2} \eta_+'^2 + {\it c.c.} 
\end{equation}
From this point  we  omit the primes for coordinates and order parameters
and take $ \theta $ as the angle of
the normal vector $ \mbox{\boldmath $ n $} $ relative to the crystal $
x $-axis. Thus, the $x(y)$-coordinate is now always parallel
(perpendicular) to the normal vector. In
this representation the $\theta$  dependence clearly does not appear, if the
system is cylindrically symmetric,  i.e., $ \nu = 0 $. 

\subsection{Structure of the domain wall}

We now turn to the domain wall and consider the
situation in which for $ x \to \pm \infty $ the superconducting states $
\eta_{\pm} $ is realized and around $ x = 0 $ a smooth change between
\begin{figure}
 \epsfxsize = 6 cm 
 \centerline{\epsfbox{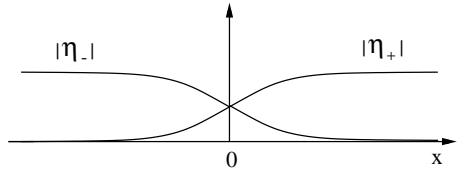}}
\caption{Schematic behavior of the order parameter components at the
 domain wall.}
\label{fig:1}
\end{figure}
the two states occurs within a finite length scale (Fig. \ref{fig:1}).
This situation is similar to a 
Josephson junction between two
superconductors with the state $ \eta_+ $ and $ \eta_- $,
respectively. Indeed the phase difference between the two states plays 
a similarly important role for the physical properties of the domain
wall. The structure of the domain wall is obtained by
a variational minimization of the free energy $ {\cal F} $. For the sake 
of transparency we simplify the treatment by introducing the 
following approximate variational ansatz for the order parameter:

\begin{equation}
\eta_+ = \eta_0 e^{i \phi_+} \cos \chi \qquad {\rm and} \qquad \eta_-
= \eta_0 e^{i \phi_-} \sin \chi .
\label{OP}
\end{equation}
This leads to the boundary conditions
\begin{equation}
\chi = \left\{ \begin{array}{ll} 0 & \qquad x \to + \infty , \\  & \\
\displaystyle \frac{\pi}{2} & \qquad x \to - \infty \end{array}. \right.
\end{equation} 
We assume spatial dependence only for $ \chi $ and leave the phase difference 
between the two sides $ \alpha = \phi_+ - \phi_- $ constant. 
The effective free energy becomes 

\begin{equation} \begin{array}{ll}
\tilde{\cal F} = \int d^3r & [ \displaystyle 
a \eta_0^2 + b \eta_0^4 + \frac{b
\eta_0^4}{2} (1 + \nu \cos(2\alpha + 4\theta)) \sin^22 \chi \\ & \\ &
\displaystyle 
+ \kappa \eta_0^2 \{ |\mbox{\boldmath $ D $} \cos \chi |^2 + |
\mbox{\boldmath $ D $}  \sin \chi |^2 +
[e^{-i\alpha} (D_- \cos \chi)^* (D_+ \sin \chi) \\ & \\ &
\displaystyle + \nu e^{-i(\alpha +
4 \theta)} (D_+ \cos \chi)^* (D_- \sin \chi)+c.c]/2 \} + \frac{1}{8\pi}
(\nabla \times \mbox{\boldmath $ A $})^2 ]. \\ &
\end{array} \end{equation}
We neglect contributions from $A_y$ 
and vary the free energy with respect to $ \chi $, 

\begin{equation} 
\partial_x^2 \chi = \frac{Q}{4} \sin 4 \chi -4S_+\gamma A_x(\partial_x \chi)
+ 2C_+(\sin 2 \chi
\partial_x^2 \chi + \cos 2 \chi [(\partial_x \chi )^2+\gamma^2A_x^2]),
\label{SG1}
\end{equation}
and $A_x$,
\begin{equation}
A_x = \frac{S_+ \partial_x \chi}{\gamma(1+C_+ \sin 2 \chi)},
\label{ax}
\end{equation}
with

\begin{eqnarray} 
Q &= \frac{1}{\xi_0^2} \{1 + \nu \cos (2\alpha + 4\theta)\}, \\
C_{\pm} & = \left [\cos \alpha \pm \nu \cos (\alpha
+ 4 \theta)\right ]/2, \nonumber \\
S_{\pm} & =\left [\sin\alpha\pm\nu\sin(\alpha+2\theta)\right ]/2,
\nonumber
\end{eqnarray} 
and $ \xi_0^2 = \kappa/ 2 b \eta_0^2  $. The third  term on the
right-hand side 
of Eq. (\ref{SG1}) and the $\sin{2\chi}$ term in the denominator of $A_x$ 
complicate the solution. 
We find, however, that the influence of these terms is small
at larger distances from the domain wall, as 
one can see in approximating $ \chi (x) \approx e^{- K x} $ in 
the limit $ x \to  + \infty $. With this ansatz, Eq. (\ref{SG1}) becomes
\begin{equation}
(1-S_+^2)K^2 e^{- K x} = Q e^{- K x} + 
O(e^{-2 K x}).
\end{equation}
Thus the second term decays much faster than the first one. Therefore, 
the width of the domain wall is largely determined by the first term, 
and the second term enters only into the domain wall energy. 

\begin{figure}
\epsfxsize = 6 cm 
\centerline{\epsfbox{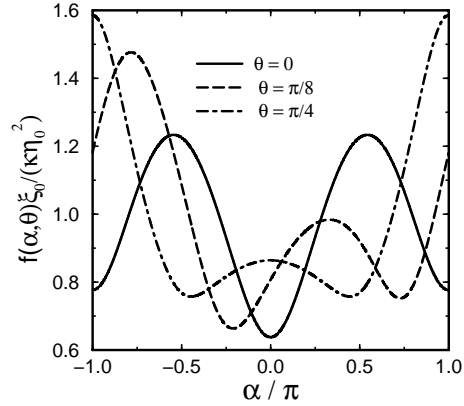}}
\caption{The domain wall energies as functions of the phase difference
 $ \alpha = \phi_+ - \phi_- $ for angles $ \theta = 0, \pi/8 $ and 
 $ \pi/4 $. The anisotropy parameter is chosen $ \nu = -0.5 $. 
}
\label{fig:2}
\end{figure}

We solve Eq. (\ref{SG1}) neglecting the third term and the $\sin{2\chi}$
term in the denominator of $A_x$, so that it takes the 
form of an ordinary Sine-Gordon equation,
\begin{equation}
\partial_x^2 \chi = \frac{\tilde{Q}}{4} \sin 4 \chi,  
\end{equation}
where $\tilde{Q}=Q/(1-S_+^2)$
for which we easily find the kink solution
\begin{equation}
\chi (x) = {\rm arctan} (e^{-x \sqrt{\tilde{Q}}}).
\label{varwavefunc}
\end{equation}
This is used to determine $A_x$ through Eq. (\ref{ax}) and both $A_x$ and 
$\chi$ are inserted into the free energy  
and integrated, so that we
obtain for the variational domain wall energy
\begin{equation}
f(\alpha, \theta) = \kappa \eta_0^2\sqrt{\tilde{Q}}
\left\{1-\frac{C_+\pi}{8}-\frac{S_+^2}{C_+}\left[
\frac{\pi}{4}-\frac{1}{\sqrt{1-C_+^2}}
{\rm arctan}\left(\frac{\sqrt{1-C_+^2}}{1+C_+}\right)\right ] \right\}
\label{kapfreen}
\end{equation}
as a function of $\alpha$ and $\theta$.

From this point we fix $ \nu < 0 $. For sufficiently large $ | \nu | $ the
domain wall energy develops two local minima as a function of the
relative phase $ \alpha $. This is illustrated in Fig. \ref{fig:2} for
angles $ \theta = 0, \pi/8 $ and $ \pi/4 $.
For the most stable orientation $ \theta =
0 $ these minima are at $ \alpha =0 $ (stable state) and $ \alpha =
\pi $ (metastable state). For $ \theta \to \pi/4 $ we find that
the stable and metastable state approach in energy and become
degenerate at exactly $ \theta = \pi/4 $.  
The presence of stable and metastable states 
is important when we discuss vortex states on the
domain wall.

\section{Vortices on the domain wall}

We now investigate local modifications of the domain wall structure which
give rise to vortices. This problem has been previously considered for
the case of the degenerate domain wall state. \cite{DOMAIN} 
Here we
extend the discussion to the general situation. 

\subsection{Fractional flux lines}

As mentioned in the Introduction, domain walls act as pinning regions
for the vortices (core pinning) due to the locally  diminished 
condensation energy. 
The resulting pinning
effect is, however, rather small, since the domain wall 
only provides a shallow attractive potential. 
In the following we would like to show that a change of the vortex
structure can lead to considerable strengthening of the pinning
potential. 

To some extent the domain wall can be viewed as a
planar weak link between two superconductors with order parameters
slightly interpenetrating, as described by Eq. (\ref{OP}).
We use this form for the order parameter to express the current
which is obtained from the variation of the free energy with respect
to the vector potential, 

\begin{equation} \begin{array}{l} \displaystyle
j_x = \frac{\kappa \gamma}{c} [2 \cos^2 \chi u_{x+} + 2 \sin^2 \chi u_{x-}
- \sin 2 \chi \{ (u_{x+} + u_{x-} ) C_+ + (u_{y+} +
u_{y-}) S_- \} + 2 (\partial_x) \chi S_+], \\ \\
\displaystyle
j_y = \frac{\kappa \gamma}{c} [2 \cos^2 \chi u_{y+} + 2 \sin^2 \chi u_{y-}
- \sin 2 \chi \{ (u_{y+} + u_{y-} ) C_+ + (u_{x+} +
u_{x-}) S_- \} + 2 (\partial_x) \chi C_-] , \\ 
\end{array} \end{equation}
with 

\begin{equation}
u_{\mu \pm} = \eta^2_0 ( \partial_\mu \phi_{\pm} - \gamma
A_\mu) .
\end{equation}
In the stable and metastable states for given $ \theta $, the current
perpendicular to the 
domain wall should vanish. If this were not the case,
current would flow everywhere in
the bulk, leading to a large energy cost. Neglecting the contributions
of $A_y$ leads to Eq. (\ref{ax}) for $A_x$, so that the variational solution 
already ensures that the current perpendicular to the wall vanishes
for $ \alpha $ of the (meta)stable state.

\begin{figure}
 \epsfxsize = 6 cm 
 \centerline{\epsfbox{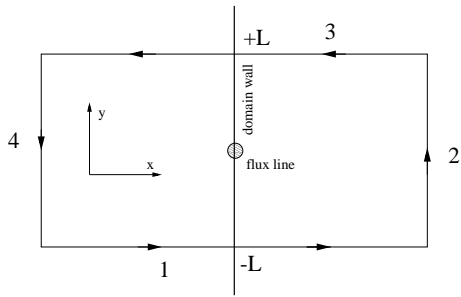}}
\caption{Rectangular path encircling the flux line. We separate the
 path into  parts 1, 2, 3 and 4 in order to calculation the
 contribution to the flux.}
\label{fig:4}
\end{figure}

We now consider vortices on the domain wall corresponding to the
winding of the order parameter phase. By analogy  to Josephson
junctions, such  a
vortex may be considered a ``kink'' of the phase difference $ \alpha
$. In the ordinary case, the size of the kink is an integer multiple of
$ 2\pi $. For the domain wall it is, however, possible to generate a
kink to a metastable state. While, in general, a $ 2\pi $-kink is
associated with the standard magnetic flux quantum $ \Phi_0 = hc/2e $, a
smaller kink between the stable and metastable state would carry a
fraction of $ \Phi_0 $ only, depending on the size of the kink and
other parameters. 
For the Josephson junction we can describe the vortex using a Sine-Gordon
equation of the phase difference containing only one length scale, the 
Josephson penetration depth. This analogy, however, is limited, because 
in the case of the domain wall two length scales have to be taken into account:
the coherence length $ \tilde{\xi}$ (length scale of the variation of
$ \alpha $) along the domain wall and the
London penetration depth $ \lambda_L $. We consider the limit $
\lambda_L \gg \tilde{\xi} $,  
ignore the detailed structure of the vortex on small length
scales ($ \sim \tilde{\xi} $) and focus mainly on its magnetic properties. 

Let us analyze now a kink of $ \alpha $ as a line defect in the domain
wall between the stable state with $ \alpha = \alpha_0 $ and the
metastable state with $ \alpha = \alpha_1 $. We encircle this line
by a wide rectangular path (see Fig. \ref{fig:4}) and calculate the
enclosed magnetic 
flux $ \Phi = \oint d{\bf s} \cdot \mbox{\boldmath $ A $} $. 
First, we consider the parts of the path which cross the
domain wall perpendicularly, 1 and 3. We choose the gauge such that $
\partial_x 
\phi_{\pm} = 0 $ along both paths, i.e., any variation of $ \phi_{\pm}
$ is restricted to the parts of the path parallel to the domain wall.
If we now use the condition $ j_x = 0 $ along this path, we obtain the
contribution 

\begin{equation}
\varphi_1 = \int^{+\infty}_{-\infty} d x A_x (x,y=-L) = \int^{\pi}_0 d
\beta \frac{S_+}{2 \gamma} \frac{1}{1 + C_+ \sin \beta} = \Phi_0
\Pi(\alpha_0, \theta)
\end{equation}
to the flux from path 1, where

\begin{equation}
\Pi(\alpha, \theta) = \frac{S_+}{2 \pi \sqrt{1 - C^2_+}} \left[ {\rm
arctan} \left(\frac{C_+}{\sqrt{1 - C^2_+}} \right) - \frac{\pi}{2}
\right] .
\end{equation}
Analogously, the contribution from path 3 is given by

\begin{equation}
\varphi_3 = \int^{-\infty}_{+\infty} d x A_x (x, y = + L ) = - \Phi_0
\Pi(\alpha_1,\theta) ,
\end{equation}
and for the paths 2 and 4 we take the variation of $ \phi_{\pm} $ into
account. With $ j_y = 0 $ sufficiently far from the domain wall, this leads to

\begin{equation} \begin{array}{ll}
\varphi_2 + \varphi_4 & =\displaystyle  \int^{+L}_{-L} d y ( A_y (x \to +
\infty,y) - A_y (x 
\to - \infty,y)) \\ & \\ & \displaystyle = \frac{1}{\gamma}
\int^{+L}_{-L} ( \partial_y \phi_+  
\partial_y  \phi_-) = \frac{\alpha_1 - \alpha_0 + 2 \pi n}{2 \pi}
\Phi_0  \\ &
\end{array}
\end{equation}
Thus, the flux for the line defect depends on the angle $ \theta $ as

\begin{equation}
\Phi(\theta) = \Phi_0 \left[\frac{\alpha_1 - \alpha_0}{2 \pi} +
\Pi(\alpha_0, \theta) - \Pi(\alpha_1, \theta) \right] + \Phi_0 n ,
\end{equation}
which determines the flux up to an integer multiple of $ \Phi_0 $. 
The possible magnetic fluxes are fractional and the smallest ones are 
smaller than $ \Phi_0 $ in magnitude. This is analogue to the 
Josephson vortices on a time-reversal symmetry breaking Josephson
junction.\cite{LAUGHLIN,KUBOKI,KUKLOV,BELZIG} In that case, however,
the two junction states separated by 
the flux line are degenerate. This is the case here only for the
special angles 
$ \theta = \pm \pi/4 $. In the special case $ \theta =0 $ the
phase differences $ \alpha_0 = 0 $ and $ \alpha_1 =\pi $ lead to $
\Phi = \pm \Phi_0 /2 $, half a flux quantum. In all other cases the
flux depends continuously on $ \theta $ and $ \nu $.

\begin{figure}
 \epsfxsize = 6 cm 
 \centerline{\epsfbox{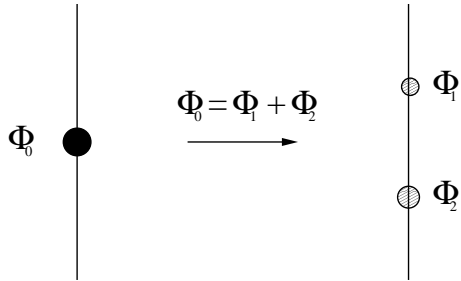}}
\caption{A standard vortex with flux $ \Phi_0 $ on the domain wall
 decays into two fractional vortices whose fluxes add up to $ \Phi_0 $.}
\label{fig:5}
\end{figure}

It is easy to see that two kinks in sequence (stable $ \to $ metastable 
$ \to $ stable) yield a total flux $ \Phi = \Phi_0 n $. Therefore, 
a conventional vortex (flux $ \Phi_0 $) may decay into two
fractional vortices on the domain wall (Fig. \ref{fig:5}). 
Through this dissociation, the magnetic field energy 
can be reduced. If we assume that the domain wall is in the
stable state, then the splitting of the vortex introduces a metastable
domain wall connecting the two fractional flux lines. Thus, the cost
of domain wall energy has to be compensated by the reduction of
magnetic energy.  

Let us compare the two energies here. The domain wall energy cost per
unit area is $ \epsilon_{dw} = f_{dw}(\alpha_1 , \theta) - f_{dw}
(\alpha_0 , \theta) $, with $ f_{dw} $ defined in Eq. (\ref{kapfreen}). 
The magnetic energy of the flux lines can be estimated by assuming
that their field distribution is not much different from that of a
standard vortex described by the London equation with the free
energy functional 
\begin{equation}
F_{L} = \int d^3r [ {\bf B}^2 + \lambda_L^2 ( \nabla \times {\bf B})^2 
]  ,
\end{equation}
where we assume that the London penetration depth is not
modified by the domain wall. 
We now consider two fractional fluxes, $ \Phi_1 , \Phi_2 >0 $, with $
\Phi_1 + \Phi_2 = \Phi_0 $ (field along the $z$-axis) at the positions 
$ {\bf r}_1 $ and $ {\bf r}_2 $, respectively. The resulting
London equation is 

\begin{equation}
B_z - \lambda_L^2 \nabla^2 B_z = \Phi_1 \delta^{(2)}({\bf r} - {\bf r}_1) +
\Phi_2 \delta^{(2)}({\bf r} - {\bf r}_2),
\end{equation}
which is taken in two dimensions, assuming homogeneity along the
$z$-axis ($ {\bf r} = (x,y) $). The solution of the equation leads to
the line energies $ \varepsilon_1 $ and $ \varepsilon_2 $ for the two
vortices and an interaction energy that depends on the distance between the
vortices. Thus
the total magnetic energy is

\begin{equation}
v(|y_2 - y_1|) = \varepsilon_1 + \varepsilon_2 +
\frac{2 \Phi_1 \Phi_2}{(4 \pi \lambda_L)^2} K_0 \left(\frac{|y_2 -
y_1|}{\lambda_L} \right), 
\end{equation}
with $ K_0 $ denoting the MacDonalds function. The line energies
contain as a lower cutoff length $ \tilde{\xi}_j
$ ($j=1,2$), which is of a magnitude similar to the coherence length,

\begin{equation}
\varepsilon_j = \frac{\Phi_j^2}{ 4 \pi \lambda_L)^2}{\rm ln} \left(
\frac{\lambda_L}{\tilde{\xi}_j} \right) .
\end{equation}
The potential for the fractional vortex at a distance $ R = |y_2 -
y_1| $ is therefore

\begin{equation}
V(R) = v(R) +  \epsilon_{dw} R - \varepsilon_0 ,
\end{equation}
where $ \varepsilon_0 $ is the line energy per unit length of a
standard vortex. The energy loss due to the
metastable state introduces a string potential between the two
vortices. The optimal distance  $R $ is chosen to minimize the potential
energy $ V(R) $, which corresponds to a pinning potential for the vortex. 
Note that this potential is only valid for $ R \gg \tilde{\xi}_j $.
Decaying into two fractional vortices leads to a very effective
pinning, since the recombination is necessary in order to
dislocate the vortex 
again from the domain wall. This occurs if the two fractional
vortices are forced to approach to $ R \sim \tilde{\xi}_j $. 

For a comparison of the energies involved, we estimate the magnetic energy
gained by the decay into fractional vortices, 

\begin{equation}
\frac{\Phi_0^2}{(4 \pi \lambda_L)^2},
\end{equation}
where the London penetration depth is $ \lambda_L^2 = 1/8\pi \kappa \eta_0^2
\gamma^2 $. The maximal string energy is obtained for $ \theta =0 $,
where 

\begin{equation}
\epsilon_{dw} R \sim \frac{\Phi_0^2}{(4 \pi \lambda_L)^2} \frac{\kappa}{8}
(1 + \nu )^{3/2} \frac{R}{\lambda_L}.
\end{equation}
In this case the two energies become comparable for $ R $ just a
fraction of $ \lambda_L $. For angles $ \theta $ closer to $ \pm
\pi/4 $, the string potential is smaller and the optimal separation is
larger. Furthermore, we find that increasing the anisotropy of the
Fermi surface, denoted by $ \nu $, tends to stabilize fractional
vortices. The stability of the fractional vortices does also depend on
temperature, because the internal structure of the domain wall can change
with lowering temperature. \cite{OGAWA} 

In this discussion we have assumed that the domain wall is 
an infinite plane with one fixed $ \theta $. In real materials this
is not the case, because domain walls are pinned at
impurities and lattices defects and, consequently, may 
change orientation by having
``corners''. 
Since the phase value $ \alpha $ for the stable domain
wall structure  depends on 
$ \theta $, a kink of $ \alpha $ should occur at every 
such corner and, according to our analysis, be accompanied by a flux
line. Similarly, one can see that the crossing of two 
domain walls introduces a 
fractional vortex along the cutting line. In general, the phase and
flux structure of a non-planar domain wall can be rather complicated.
However, we do not consider these aspects any further here.
 
\subsection{Barrier effect of a domain wall}

\begin{figure}
 \epsfxsize = 5 cm 
 \centerline{\epsfbox{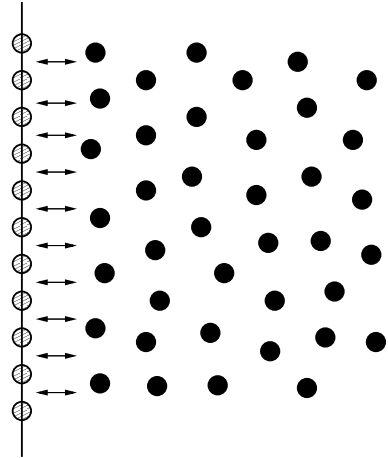}}
\caption{Schematic picture of a domain wall occupied by fractional
 flux line (shaded) which repel approaching vortices (filled circles) 
 so that they do not traverse the domain wall.}
\label{fig:6}
\end{figure}

We now turn to the situation in which the superconductor is in the mixed
phase and contains many vortices. If the conditions are appropriate, 
some of the vortices will be trapped by
the domain wall and decay into fractional vortices. 
These flux lines are strongly pinned and may form a
queue on the domain wall like the planks of a fence. Other vortices approaching
the domain wall now are repelled by this vortex fence and cannot
easily penetrate or traverse the domain wall (Fig. \ref{fig:6}). In
this way the domain wall acts as a very effective barrier for
vortices.  

The question arises under which conditions the domain wall becomes
permeable again. The vortices outside the domain wall generate a pressure 
on the fractional vortices, influencing their density and, hence, their
mutual distance. With 
increasing external vortex density $ \rho = B / \Phi_0 $ the distance $ R $
should shrink. This may be described qualitatively by adding a term 
$ a_{\rm ext} R^2 $ to the potential $ V(R) $ where $ a_{\rm ext} > 0
$ grows with increasing $ \rho $ at the domain wall. 
Hence, it is clear that a growing density of
vortices close to the domain wall forces eventually  the fractional
vortices to recombine when $ R \sim \tilde{\xi} $, the extension of
the kinks of $ \alpha $. The recombined vortices are only weakly pinned
so that the fence becomes permeable. This process needs not occur
over all the domain wall simultaneously , but may start with local leaks for
vortices to pass through. 
A simple picture of the ``barrier height'' of the domain wall
can be obtained by the following argument. The density of fractional
vortices is more or less equal to that of the vortices immediately
outside of the domain wall. Therefore, the relation for the 
spacing $ R \sim \rho^{-1/2} = (\Phi_0/B)^{-1/2} $ leads to the
barrier field $ B^* \propto \tilde{\xi}^{-2} $. The temperature
dependence of $ \tilde{\xi} $ close to the transition temperature
(the second transition in the case of a system with double transition)
is essentially proportional to $ |T-T_c|^{-1/2} $ so that the barrier
field is growing as the temperature is lowered $ B^* \propto | T -T_c|
$.  

With these properties,
the presence of domain walls should have considerable
influence on the flux motion in the lower range of magnetic fields in
the mixed phase. In particular this is true for phenomena like
flux creep and hysteresis behavior of the mixed phase.
We first discuss the influence of the barrier effect on the flux creep 
which can be observed in the relaxation of remnant magnetization after a
magnetic field has been applied for some time to a superconductor and then
switched off. The flux relaxation behavior has been investigated in
much detail by the group of Mota for the heavy Fermion superconductors
UPt$_3$ and U$_{1-x}$Th$_x$Be$_{13}$ and for
Sr$_2$RuO$_4$. \cite{AMANN,LT,MOTA} The first two 
superconductors exhibit double transitions, where 
the low-temperature phase in each case is very likely
time-reversal symmetry breaking and should therefore provide the
conditions for domain walls as described above. Mota and coworkers
observed that the initial flux relaxation rate drops drastically below
the second superconducting transition. Hence, many vortices generating
the remnant magnetization find it apparently more difficult to escape 
from the sample in the low-temperature phase. While initially almost
no remnant flux leaves,  
the relaxation recovers after a longer waiting time. 
The absence of flux relaxation can be understood as the
barrier effect of the domain wall. In the low-temperature phase, 
domains appear covering the sample with many domain walls. Vortices of 
the remnant magnetization are fenced in by these walls, and only a
small portion of the total flux (mainly that close to surface not impeded
by barriers) can move out when the external field is turned
off. However, the encircled vortices press the 
domain walls which then may move slowly, and, finally, new pathways can open
for the flux lines to leave the sample. Thus, the later appearance of flux
decay can be attributed to slow domain wall motion. In the
high-temperature phase domain walls are absent leading to 
flux creep as usual. 

The condition for this behavior is the existence of
a large number of domain walls 
throughout the sample which are pinned at impurities and lattice
defects such that they do not move too easily. Domains are believed to 
nucleate randomly at the second superconducting transition if there is 
no bias for one type of domain. If the superconducting
state breaks time-reversal symmetry, an external magnetic field could
provide a bias. Therefore we may expect that if the superconductor is
cooled in a magnetic field (at least for the time-reversal symmetry breaking
transition), the distribution of domains should be unbalanced in favor
of one type reducing the number of domain walls. Then, the flux
creep, previously impeded by domain walls, would be more
ordinary. \cite{MOTADIS}
A similar picture arises also from the measurement of flux motion in 
the low-temperature phase of U$_{1-x}$Th$_x$Be$_{13}$ by 
Zieve et al.. \cite{ZIEVE} 
The samples which are field cooled and
zero-field cooled exhibit a difference with regard to the motion of flux. 
In the former case, flux lines move more easily through the sample than in
the latter, since in the zero-field cooled case, the presence of domain walls
would also here impede the flux flow.
Thus, there is a clear difference between the field- and
zero-field cooled situation. The effect continuously grows below the
second transition, indicating a gradual increase of the barrier
strength as suggested above. This is an
aspect that also appears in the hysteresis experiment discussed in
the next section. \cite{SHUNG}

\section{Intrinsic noise in the hysteresis}

The hysteresis of the magnetization in the mixed phase provides one way to 
measure the critical current. The basic features of the hysteresis
cycle are described well by the critical vortex state models among
which the Bean model is the most simple representation. \cite{BEAN} 
In this model the vortices are described as a magnetic flux 
density. We omit here a detailed introduction, since the basic
properties of the Bean model can be found in many textbooks. \cite{TINKHAM} 

\begin{figure}
 \epsfxsize = 12 cm 
 \centerline{\epsfbox{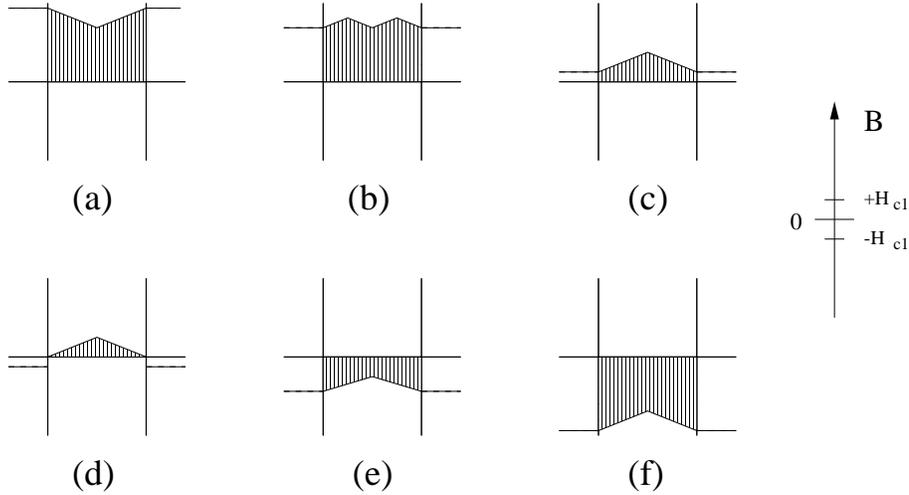}}
\caption{Standard Bean profile of the flux density 
for the magnetization process in the
 hysteresis of a superconductor in the mixed state. The magnetization
 $ M(H) $ is  the integral of flux density in the sample.}	
\label{fig:7}
\end{figure}

For the following analysis  we use the model of a superconducting slab
which has a finite  
width along the $ x $-axis and is infinite along the other two
axes. Let us first study the standard behavior of the hysteresis,
including the surface barrier effect of the lower critical field (or
alternatively the Bean-Livingston barrier). We start by assuming
\begin{figure}
 \epsfxsize = 7 cm 
 \centerline{\epsfbox{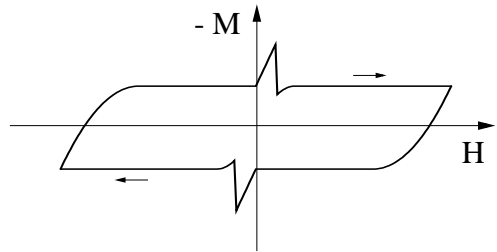}}
\caption{Standard hysteresis in the mixed state of a superconductor
 based on the Bean critical state model.}
\label{fig:7a}
\end{figure}
that the external field was first raised to the maximum
value $ H_m $. Then the field distribution is similar to that in 
Fig. \ref{fig:7}a). 
Now we turn the field off gradually and reverse it until we reach $ - H_m 
$ (Fig. \ref{fig:7}f). On the way the field distribution passes through the
distributions shown in Fig. \ref{fig:7}b)-e). It is important that in
order to introduce reversed vortices into the system, the external
field must exceed the lower critical field. Hence, in the range
between $ H_{\rm ext} = 0 $ 
and $ - H_{c1} $, the internal flux distribution is unchanged, and at $ -
H_{c1} $ the magnetization jumps abruptly. After reaching $ H_{\rm ext}= - 
H_m $, the field is again gradually turned to zero and further to
$ H_m $, so that the hysteresis cycle closes. The magnetization $ M $
as a function of  $ H $  
corresponds to the integrated flux density in the Bean critical state. 
The slope of the field distribution corresponds to the critical
current ($ J_c = (c/4\pi) \partial_x B_z $) and is proportional to the
width of the hysteresis loop (Fig. \ref{fig:7a}).

\begin{figure}
 \epsfxsize = 12 cm 
 \centerline{\epsfbox{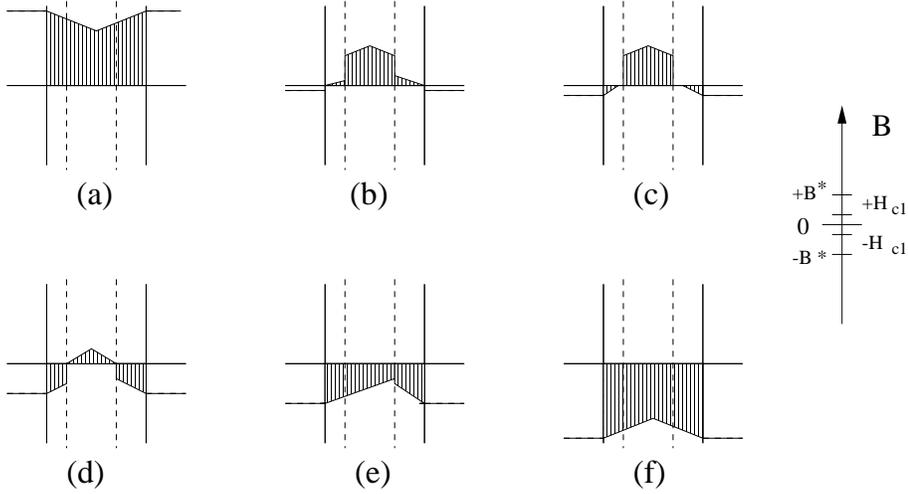}}
\caption{Bean profile of a superconducting slab with two domain walls
 in a magnetic field. The critical field and the barrier
 field are indicated on the axis on the right hand side.}
\label{fig:10}
\end{figure}

If we now introduce domain walls as additional barriers within the
sample, new structures appear in the hysteresis cycle. The domain wall
is characterized by a specific barrier field (height) $ B^*
$;  i.e., if the local field at the domain wall exceeds $ B^* $, the domain
wall is permeable to vortices. 
For simplicity we consider the case of two domain walls which 
fence in some region inside the slab. We start again from the
situation in which the external field is at the value $ H_m$, so that  
everywhere in the sample the local field value is  larger than  
$ B^* $, and the
Bean profile has developed everywhere with a slope corresponding to
the critical current due to usual pinning (Fig. \ref{fig:10}a). Now, we
gradually reduce the external field. As we reach zero, we find that the
domain walls have trapped some flux; that is,  more flux than
usual is remaining in
the sample (Fig. \ref{fig:10}b). If we now turn to negative fields,
then initially at $ - H_{c1} $, reversed flux lines enter and
annihilate the 
flux close to surface (Fig. \ref{fig:10}c). With decreasing 
external field, the reversed flux lines reach the domain walls and
annihilate vortices on the domain wall, which are in turn immediately 
replaced by new positive flux lines from the interior of the fenced in
region.  This process continues until the density of positive vortices
on the domain wall is essentially zero and leads to a sharp increase in the
magnetization similar to that at the lower critical field
(Fig. \ref{fig:10}d).  
Note that the negative vortices from the outside cannot enter the
inner region yet, since at the end of this process the domain wall
is occupied by reversed (negative) fractional vortices, which now provide
a barrier. The flux density must be enhanced from outside, and
the domain wall only becomes permeable when the local field of the
negative vortices has reached $ - B^* $ (Fig. \ref{fig:10}e). Then,
negative vortices can pass 
freely through the domain wall and annihilate the remaining positive
vortices on the other side. This leads to a further avalanche-like
rise in the magnetization. Then the external field is lowered until we 
reach $ - H_m $ (Fig. \ref{fig:10}f). The analogous continuation for
increasing field finally leads to a closed hysteresis cycle. 

The additional sharp avalanche-like structures in the magnetization
depend strongly on the position of the domain walls. As mentioned
earlier, the domain walls 
\begin{figure}
 \epsfxsize = 8 cm 
 \centerline{\epsfbox{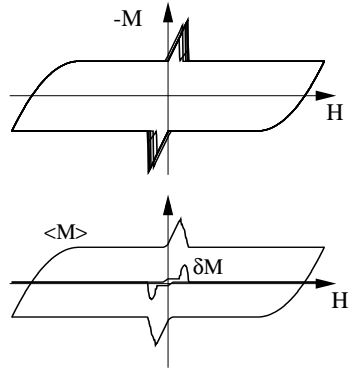}}
\caption{Hysteresis of a superconducting slab containing domain walls
 as barriers. Upper figure: Several cycles with varying location of
the domain wall. Lower figure: Average of several cycles and standard
deviation, which possess maxima around the region of $ \pm H_{c1} $.}
\label{fig:8}
\end{figure}  
are not rigidly fixed, but only pinned, and can change
their position. During several cycles, the domain walls may be
located at different places from cycle to cycle due to pressure from the 
vortices. Hence, the hysteresis cycles do not retrace themselves 
completely, but rather have strong deviations in the field range where the
avalanche effects occur, i.e., close to the lower
critical field, $ \pm H_{c1} $. Taking the average 
over several cycles, the standard   
deviation, $ \delta M = \langle (M(H) - \langle M(H) \rangle )^2 \rangle
$ reveals a strong ``noise'' signal, as shown in Fig. \ref{fig:8} ($
\langle ... \rangle $ denotes the average over many cycles).
This noise is intrinsic
to the superconducting state with internal barriers. It is easy to 
see that the magnitude of this noise depends on the barrier height
given by $ B^* $, and it should disappear, if $ B^* $ goes to zero
(the case of a completely permeable domain wall):
\begin{equation}
\delta M \propto B^* \delta x,
\label{standdev}
\end{equation} 
where $ \delta x $ is the standard deviation of the domain wall
positions, only weakly dependent on temperature. 

We may compare this behavior now to the experiments done on UPt$_3$ by 
Shung et al.. \cite{SHUNG} The features of the noise in the hysteresis
are qualitatively very similar to those obtained in our simple
simulation.  In the high-temperature phase, the noise in the
hysteresis cycle is small and featureless. However, simultaneously
with the onset of 
the second transition, the intrinsic noise appears
continuously and grows essentially 
linearly with $ | T - T_{c2} | $ as we expect
from Eq. (\ref{standdev}).  
Within the cycle the noise appears in the same field range and with a
similar structure, as in our simulation, i.e. around $ H_{c1} $. This
suggests that the continuous increase of
the noise is very likely connected with the broken time-reversal
symmetry below the second transition where the degeneracy of the
superconducting phase allows for domain
formation. The barrier height introduced by these domain walls
increases as the order parameter of the low-temperature phase grows. 
A similar picture arises also from the measurement of flux motion in 
the low-temperature phase of U$_{1-x}$Th$_x$Be$_{13}$ by 
Zieve et al., \cite{ZIEVE} as mentioned above. 

\section{Conclusion}

We have shown that domain walls in time-reversal symmetry-breaking
superconductors can play an important role for  the motion of flux
lines. Under certain conditions, domain walls can accommodate flux
lines whose flux is a fraction of the standard flux
quantum. On such domain walls, conventional vortices can decay into two 
fractional vortices. If such fractional vortices line up,  
the domain
wall becomes a very efficient barrier to vortices. 

Evidence for this property of the domain wall is given by the absence
of flux creep or in the intrinsic noise of the hysteresis
cycle. \cite{AMANN,LT,MOTA,SHUNG,ZIEVE} In
the case of UPt$_3$ and U$_{1-x}$Th$_x$Be$_{13}$ experiments reveal a
clear connection between the onset 
of the time-reversal symmetry-breaking in the low-temperature phase and the
appearance of a new flux trapping mechanism. In Sr$_2$RuO$_4$ the drop
of the creep rate is not connected with the onset of the time-reversal
symmetry breaking phase. Rather the drop around 50 mK may be
associated with a ``transition'' of the domain wall states. This may
be associated with the multi-band nature of the superconducting state, 
which can lead to the temperature dependence of parameters like the
anisotropy factor $ \nu $, since the relative contribution to
superconductivity by the different bands certainly varies with
temperature. This case needs definitely further consideration. 
The barrier
effect of domain walls does not, in general, influence the magnitude
of the critical current $ J_c $, which is related to ordinary
impurity-induced pinning effects. Although the initial flux relaxation
is vanishing in the 
low-temperature phase, $ J_c $ remains finite and does not show any
anomalous temperature dependence, indicating that the
phenomena are not connected with a change in the pinning of individual
vortices.
We may take the observation of such flux flow phenomena as indirect
evidence 
for the presence of domain walls. To this time in none of the above
mentioned  
superconductors has the direct observation of the domain walls been
reported. Clearly, the discovery of lined-up fractional vortices could
be one direct experimental verification. Another proposal is based on
the modification of the local quasiparticle density of states in the
domain wall which may be observable by scanning tunneling microscopy.
\cite{MATSUMOTO} This paper presents a plausible explanation for the
basic properties seen in the experiments of three unconventional 
superconductors. Nevertheless, several details in the
experimental data, which may be typical for the phenomenon or specific
to each system, remain to be discussed in future. 

\section*{Acknowledgements}

We would like to thank A. Amann, V. B. Geskenbein, R. Joynt,
T. M. Rice, T. F. Rosenbaum and E. Shung, and especially A.-C. Mota for many 
helpful and inspiring discussions. This project has been financially
supported by a Grant-in-Aid of the Japanese Ministry of Education,
Science, Sports and Culture. DFA acknowledges the financial support of NSF
cooperative grant No. DMR-9527035 and the State of Florida and also  
thanks the Aspen Center for Physics, where  
this work was completed.


\begin{thebibliography}{99}
\bibitem{BLATTER} G. Blatter, M. V. Feigel'man, V. B. Geshkenbein,
A.I. Larkin and V. M. Vinokur, Rev. Mod. Phys. {\bf 66} (1994), 1125.
\bibitem{AMANN} A. Amann, A.C. Mota, M. B. Maple and H. v.L\"ohneysen,
Phys. Rev. {\bf B57} (1998), 3640.
\bibitem{LT} E. Dumont, A.-C. Mota and J.L. Smith, {\it Proceeding of
LT22}, (1999). 
\bibitem{MOTA} A.C. Mota, E. Dumont, A. Amann and Y. Maeno, Physica B
{\bf 259-261} (1999), 934. 
\bibitem{SHUNG} E. Shung, T. F. Rosenbaum and M. Sigrist,
Phys. Rev. Lett. {\bf 80} (1998), 1078.
\bibitem{ANACELIA} A.-C. Mota, G. Juri, P. Visani, A. Pollini,
T. Teruzzi and K. Aupke, Physcia C {\bf 185-189} (1991), 343.
\bibitem{KIM} P. W. Anderson and Y. B. Kim, Rev. Mod. Phys. {\bf 36}
(1964), 39.
\bibitem{TINKHAM} M. Tinkham, {\it Introduction to Superconductivity},
(McGraw-Hill, 1996).
\bibitem{GESH} V. B. Geshkenbein and A. I. Larkin,
Zh. Eksp. Teor. Fiz. {\bf 95} (1989), 1108 [Sov. Phys.-JETP {\bf 68}
(1989), 639].
\bibitem{BEAN} C. P. Bean, Rev. Mod. Phys. {\bf 36} (1964), 31.
\bibitem{REVIEW} M. Sigrist and K. Ueda, Rev. Mod. Phys. {\bf 63} (1991), 
239.
\bibitem{VOLOVIK1} G. E. Volovik and L. P. Gor'kov, Pis'ma
Zh. Eksp. Teor. Fiz. {\bf 39} (1984), 550 [JETP Lett. {\bf 39}
(1984), 674].
\bibitem{DOMAIN} M. Sigrist, T. M. Rice and K. Ueda,
Phys. Rev. Lett. {\bf 63} (1989), 1727.
\bibitem{MATSUMOTO} M. Matsumoto and M. Sigrist,
J. Phys. Soc. Jpn. {\bf 68} (1999), 994.
\bibitem{VOLOVIK2} M. T. Heinil\"a and G. E. Volovik, Physica {\bf B210} 
(1995), 300.
\bibitem{LAUGHLIN} M. Sigrist, D. B. Bailey and R. B. Laughlin,
Phys. Rev. Lett. {\bf 74} (1995), 3249.
\bibitem{KUBOKI} K. Kuboki and M. Sigrist, J. Phys. Soc. Jpn. {\bf 65} 
(1996), 361.
\bibitem{KUKLOV} A. B. Kuklov, Phys. Rev. {\bf B52} (1995), R7002.
\bibitem{BELZIG} W. Belzig, C. Bruder and M. Sigrist,
Phys. Rev. Lett. {\bf 80} (1998), 4285.
\bibitem{OGAWA} M. Sigrist, N. Ogawa and K. Ueda,
J.~Phys.~Soc.~Japan~60 (1991), 2341.
\bibitem{MOTADIS} A.-C. Mota, private communication.
\bibitem{ZIEVE} R. J. Zieve, T. F. Rosenbaum, J. S. Kim, G. R. Stewart and 
M. Sigrist, Phys. Rev. {\bf B51} (1995), 12041.

\end{thebibliography}
\end{document}